\documentstyle[12pt,epsf]{article}
\newcommand{\be}{\begin{equation}}
\newcommand{\ee}{\end{equation}}
\newcommand{\bea}{\begin{eqnarray}}
\newcommand{\eea}{\end{eqnarray}}
\newcommand{\bt}{\begin{tabular}}
\newcommand{\et}{\end{tabular}}
\newcommand{\ba}{\begin{array}}
\newcommand{\ea}{\end{array}}
\newcommand{\ov}{\overline}

\begin{document}
\setcounter{page}{0}
\thispagestyle{empty}
\baselineskip=20pt

\hfill{
\begin{tabular}{l}
DSF$-$95/28\\
INFN$-$NA$-$IV$-$95/28
\end{tabular}}

\bigskip\bigskip

\begin{center}
\begin{huge}
{\bf Neutrino propagation in a medium with a magnetic field}
\end{huge}
\end{center}

\vspace{2cm}

\begin{center}
{\Large
Salvatore Esposito$^{1}$ and Geny Capone$^{2}$\\}
\end{center}

\vspace{0.5truecm}

\normalsize
\begin{center}
{\it
\noindent
$^{1}$Istituto Nazionale di Fisica Nucleare,
Sezione di Napoli,
Mostra d'Oltremare Pad. 19-20, I-80125 Napoli Italy\\
E-mail address: sesposito@na.infn.it\\
\bigskip
$^{2}$Universitaire Instelling Antwerpen, Department Natuurkunde,
Universiteitsplein 1, \\
B-2610 Antwerpen Belgium\\
E-mail address: geny@nats.uia.ac.be}
\end{center}

\vspace{3truecm}


\begin{abstract}
We study the properties of neutrinos propagating in an isotropic
magnetized medium in the two physical approximations of degenerate
Fermi gas and classical plasma.
The dispersion relation shows that, for peculiar
configurations of the magnetic field, neutrinos can propagate freely
as in vacuum, also for very large density; this result can be very
important in the study of supernova evolution. For mixed neutrinos,
the presence of a magnetic field can alter significatively MSW
oscillations, and for particular configurations of the field the
resonance condition no longer occurs. Furthermore,
on the contrary to that happens in non-magnetized media,
spatial dispersion arises and neutrino trajectory can
be in principle deviated; however a simple estimate shows that this
deviation is not detectable.
\end{abstract}

\newpage

\section{Introduction.}

\indent

When a neutrino passes through a medium, the interaction
with the particles in the background gives
rise to modification of the properties of the neutrino itself. For example,
even if neutrinos are exactly massless, in a medium they can acquire an
effective mass \cite{1}  and also an effective electromagnetic coupling
\cite{2}. The most famous effect for massive and non degenerate
neutrinos is the MSW effect \cite{3}: if the resonance condition
is fulfilled, for neutrinos (created by weak interactions
with a definite flavour) propagating in matter, the probability for
transition to another flavour can be appreciable different from zero also
for very small mixing angles. The solution of the solar neutrino
problem \cite{SNP} in terms of this effect \cite{4} is a well known tool.

The covariant formalism that enables one to consider the interactions
of a neutrino
passing through a medium is the Finite Temperature and
Density Quantum Field Theory \cite{FTFT} (throughout this paper we use
the real-time formulation \cite{5} of this theory, in which the
Feynman rules for the vertices are identical to the corresponding ones
in the vacuum). The effect of the temperature and of the density is taken
into account in the expressions of the free particle propagators.
For fermions and bosons we have respectively \cite{5}:
\bea
S_F (P) \; & = & \; \left( \not{P} \, + \, m \right) \, \left[
\frac{1}{P^2 \, - \, m^2} \; + \; i \, \Gamma_F (P) \right]
\label{1.1}  \\
D_{\mu \nu} (P) \; & = & \; - \, g_{\mu \nu} \, \left[
\frac{1}{P^2 \, - \, m^2} \; - \; i \, \Gamma_B (P) \right]
\label{1.2}
\eea
where
\be
\Gamma (P) \; = \; 2 \pi \, \delta (P^2 \, - \, m^2) \, \left[
\theta (P \cdot u) \, n(P) \; + \; \theta (- \, P \cdot u) \,
\bar{n}(P) \right]
\label{1.3}
\ee
and
\be
n_F (P) \; = \; \frac{1}{e^{\beta ( |P \cdot u| \, - \, \mu )} \; + \; 1}
\;\;\;\;\;\;\;\;\;\;\;\;\;\;\;\;
n_B (P) \; = \; \frac{1}{e^{\beta ( |P \cdot u| \, - \, \mu )} \; - \; 1}
\label{1.4}
\ee
are the Fermi-Dirac and Bose-Einstein distribution functions
($n_{\bar F}$ and $n_{\bar B}$ are the distribution functions for the
antiparticles). Note that in a medium another 4-vector must be
considered: the medium 4-velocity $u_{\mu}$. We restrict ourselves to
temperature which are low compared with the $W^{\pm}$ and $Z^0$
masses, so that we don't consider the temperature-dependent term
$\Gamma_B (P)$ in $D_{\mu \nu} (P)$.

In the present work we want to study the propagation of a neutrino in
a magnetized medium. An external  magnetic field forces the background
particles to describe spiral-like trajectories along the force lines,
so that it influences deeply the behaviour of the medium.
Consequently, the interaction of neutrinos with the modified
background particles modifies neutrino properties both in comparison
with the vacuum ones and in comparison with their properties in a
non-magnetized medium.

The medium we assume it consists of an electron gas, and a positive
charge density is spread homogeneously throughout in order to mantain
the charge neutrality (jellium model). Although our following analysis
works for a degenerate gas as well as for a non-degenerate gas (wheater
or not it is relativistic), when the calculations are carried
explicitely we discuss the specific cases of an electron gas in
condensed matter and in plasmas. For the former, we shall make the
simplification of a gas of free charged ``effective'' fermions (the
electrons dressed by the interaction with ions), uninfluenced by the
mutual electrostatic forces. In the latter case we stress the
relevance of the long-range electron-electron interaction, that gives
rise to a large repertoire of collective behaviour. By crossing from
the temperature range of a degenerate gas ($T \ll T_F$ where $T_F$ is
the Fermi temperature) to much higher temperatures ($T \gg T_F$), the
electron gas can be described in terms of non-quantum mechanics. The
interest in considering a plasma as medium allows us to debate
possible implications of our results in different physical contexts,
i.e. astrophysical field.

In the following section we evaluate the neutrino self-energy in a
magnetized medium in the two physical approximation of a degenerate
gas and of a classical plasma. The results here obtained are used to
study the dispersion relation in section 3 and spatial dispersion in
section 4. The modifications on neutrino matter oscillations are,
instead, analyzed in section 5. Finally, in section 6 some possible
applications of the obtained results are picked out.

\section{Neutrino self-energy in a medium with an external static
         magnetic field.}

\indent

The propagation of a neutrino, moving along the z-axis with 4-momentum
$K_{\mu}=(\omega,0,0,k)$, is described by the Dirac equation, that in
momentum space is
\be
\left( \not{K} \, - \, m \, - \, \Sigma \right) \, \psi \; = \; 0
\label{2.1}
\ee
where $\Sigma (K)$ is the neutrino self-energy. \footnote{For the sake
of simplicity, here we consider a neutrino with a Dirac mass and zero
mixing angle; the extension to non-zero mixing angles will be
discussed in section 5.}
For propagation in a non magnetized medium \cite{NR}, the one-loop
relevant contribution of the electron background to the self-energy is
given by the diagrams in fig.1,2:
\bea
- i \Sigma_W \, & = & \, \int \frac{d^4 P}{(2 \pi)^4} \,
\left( \frac{- i g}
{2 \sqrt{2}} \, \gamma_{\mu} \left( 1 - \gamma_5 \right) \right) \, i
S_F(P) \, \left( \frac{- i g}{2 \sqrt{2}} \, \gamma_{\nu}
\left( 1 - \gamma_5 \right) \right) \, i D_{W}^{\mu \nu} (K-P)
\label{2.2} \\
\vspace{1cm}
- i \Sigma_Z \, & = & \, \left( \frac{- i g}{4 cos \theta_W} \right) \,
\gamma_{\mu} \left( 1 - \gamma_5 \right) \, i D_{Z}^{\mu \nu}
\int \frac{d^4 P}{(2 \pi)^4} \, (-1) \cdot \nonumber  \\
&  & \cdot Tr \left[
\left( \frac{i g}{4 cos \theta_W} \right) \, \gamma_{\nu} \left(
1 - 4 sin^{2} \theta_{W} - \gamma_5 \right) \, i S_F(P) \right]
\label{2.3}
\eea
where $\Sigma_W$ contributes only to $\nu_e$ self-energy, while
$\Sigma_Z$ contributes also to that of $\nu_{\mu}, \nu_{\tau}$. The
fermion propagator $S_F(P)$ is taken by eq. (\ref{1.1}).

In general, the self-energy can be write as the sum of a
temperature-independent part and a temperature-dependent term; we are
interested on the effects that the medium can have on neutrino
propagation, and then we consider only the temperature-dependent term
$\Sigma(T)$. Furthermore, $\Sigma(T)$ will be complex, but we ignore
absorbitive effects, and consider only the real part of the
self-energy. Then, we can write
\be
\cal{R}e \Sigma \; = \; \frac{1 + \gamma_5}{2} \, \cal{R}e
\widetilde{\Sigma} \,
\frac{1 - \gamma_5}{2}
\label{2.4}
\ee
with
\be
\cal{R}e \widetilde{\Sigma_W} \; = \; - g^2 \, \int \frac{d^4 P}
{(2 \pi)^4} \, \frac{\Gamma_{F} (P)}{(K-P)^2 - M_W^2} \, \not{P}
\label{2.5}
\ee

\be
\cal{R}e \widetilde{\Sigma_Z} \; = \; - \frac{g^2}{2 M_Z^2 \,
cos^2 \theta_W} \,
\left( 1 - 4 sen^2 \theta_W \right) \, \int \frac{d^4 P}{(2 \pi)^4} \,
\Gamma_F(P) \not{P}
\label{2.6}
\ee
{}From Lorentz invariance, we have \cite{NR}
\be
- \cal{R}e \widetilde{\Sigma} \; = \; a_L \not{K} \, + \, b_L \not{u}
\label{2.7}
\ee
where, in the rest frame of the medium ($u_{\mu} = (1, \vec{0})$), the
coefficients $a_L,b_L$ are given by
\bea
a_L \; & = & \; \frac{1}{k^2} T_K \, - \, \frac{\omega}{k^2} T_u
\label{2.8} \\
\vspace{1cm}
b_L \; & = & \; \frac{\omega^2 - k^2}{k^2} T_u \, - \,
\frac{\omega}{k^2} T_K
\label{2.9}
\eea
where $T_K \, = \, \frac{1}{4} Tr (\not{K} \cal{R}e
\widetilde{\Sigma})$ and
$T_u \, = \, \frac{1}{4} Tr (\not{u} \cal{R}e \widetilde{\Sigma})$.

In general, $a_L,b_L$ are scalar functions of the invariants $x=K
\cdot u = \omega$ and $y=K^2 = \omega^2 - k^2$; from (\ref{2.8})
and (\ref{2.9}) let us incidentally note that the coefficients $a_L,b_L$
satisfy the following differential equations:
\be
\frac{\partial b_L}{\partial y} \, + \, x \, \frac{\partial
a_L}{\partial y} \, + \, \frac{\partial T_u}{\partial y} \; = \; 0
\label{2.11}
\ee

\be
\frac{\partial b_L}{\partial x} \, + \, x \, \frac{\partial
a_L}{\partial x} \, + \, a_L \, + \, \frac{\partial T_u}
{\partial x} \; = \; 0
\label{2.12}
\ee
At first order in $G_F$, in (\ref{2.5}),(\ref{2.6}) the $K$-dependence
vanishes, and then $a_L \simeq 0$. For $b_L$, the charged current
contribution to $\nu_e$ self-energy is \cite{NR}
\be
b_L^W \; \simeq \; - \sqrt{2} G_F ( N_e - N_{\bar e} )
\label{2.13}
\ee
where $N_e$($N_{\bar e}$) is the electron (positron) number density.
For $\nu_e$, $\nu_{\mu}$, $\nu_{\tau}$ the neutral current
contributions come from electron, proton and neutron background.
Because the coupling to the $Z^0$ of the proton is opposite to
that of the electron , for neutral media these two contributions
cancel themselves \cite{NR} and one remains only with the neutron
contribution:
\be
b_L^Z \; \simeq \; \frac{G_F}{\sqrt{2}} ( N_n - N_{\bar n} )
\label{2.14}
\ee
Then, in normal media (in which there are no antiparticles) we have
\be
b_L \; \simeq \; - \sqrt{2} G_F ( N_e - \frac{1}{2} N_n )
\label{2.14a}
\ee
for $\nu_e$, and
\be
b_L \; \simeq \; \frac{G_F}{\sqrt{2}} N_n
\label{2.14b}
\ee
for $\nu_{\mu}$,$\nu_{\tau}$.

Let us introduce now an external static magnetic field. We want to
study coherent effects of electrons (and protons) in the background on
neutrino propagation, so that the interaction region of the space in
which we are interested must be microscopically large; however, this
region is macroscopically small, and then we assume that the applied
field is also uniform in space.

Here we don't consider the possible intrinsic neutrino magnetic moment
which, in the standard model, is non-zero for massive neutrinos, but
however extremely small \cite{6}:
\be
\mu_{\nu} \; \simeq \; \frac{3 e G_F}{8 \pi^2 \sqrt{2}} \, m_{\nu}
\; \simeq \; 3 \cdot 10^{-19} \, \left( \frac{m_{\nu}}{1 eV} \right)
\mu_B
\label{2.15}
\ee
where $\mu_B \, \simeq \, 3 \cdot 10^{-7} eV^{-1}$ is the Bohr
magneton. The interaction of the neutrino with the magnetic field
proceeds only through the interaction with the electrons and protons
in the medium interacting with the field.

The extra-contributions to the neutrino self-energy come from the
diagrams in fig. 3,4: for the electron background we have:
\bea
i \Sigma_W^B \; & = & \; \lim_{\vec{q} \rightarrow 0}
\int \frac{d^4 P}{(2 \pi)^4} \, \left( \frac{- i g}
{2 \sqrt{2}} \, \gamma_{\mu} \left( 1 - \gamma_5 \right) \right) \, i
S_F(P) \, \left( i e \gamma_{\alpha} \right) \, i S_F(P + Q) \cdot
\nonumber \\
&  & \cdot \left( \frac{- i g}{2 \sqrt{2}} \, \gamma_{\nu}
\left( 1 - \gamma_5 \right) \right) \, i D_{W}^{\mu \nu} (K-P) \,
A^{\alpha}
\label{2.16} \\
\vspace{1cm}
i \Sigma_Z^B \; & = & \; \lim_{\vec{q} \rightarrow 0}
\left( \frac{- i g}{4 cos \theta_W} \right) \,
\gamma_{\mu} \left( 1 - \gamma_5 \right) \, i D_{Z}^{\mu \nu} (Q)
\int \frac{d^4 P}{(2 \pi)^4} \, (-1) \cdot \nonumber \\
&  & \cdot Tr \left[
\left( \frac{i g}{4 cos \theta_W} \right) \, \gamma_{\nu} \left(
1 - 4 sin^{2} \theta_{W} - \gamma_5 \right) \, i S_F(P+Q)
\left( i e \gamma_{\alpha} \right) i S_F(P) \right] A^{\alpha}
\label{2.17}
\eea
(in the static limit, $Q_{\mu} \, = \, (0, \, \vec{q} \rightarrow
0)$). By taking into account only the temperature-dependent real part
of the self-energy, with notation similar to (\ref{2.4}), at lowest
order in $G_F$ we get (see appendix for calculation details)
\bea
\cal{R}e \widetilde{\Sigma_W^B} & = & \lim_{\vec{q} \rightarrow 0}
\frac{8 e G_F}{\sqrt{2}} \, \int \frac{d^4 P}{(2 \pi)^4} \,
\left[ \frac{\Gamma_{F} (P)}{(P + Q)^2 - m_e^2} \, + \,
\frac{\Gamma_{F} (P+Q)}{P^2 - m_e^2} \right]
\left( T_{\alpha \mu}^{S} \, + \, i T_{\alpha \mu}^{A} \right) \,
A^{\alpha} \gamma^{\mu}
\label{2.18}  \\
\vspace{1cm}
\cal{R}e \widetilde{\Sigma_W^Z} & = & \lim_{\vec{q} \rightarrow 0}
\left( - \frac{4 e G_F}{\sqrt{2}} \right)\, \int \frac{d^4 P}{(2 \pi)^4} \,
\left[ \frac{\Gamma_{F} (P)}{(P + Q)^2 - m_e^2} \, + \,
\frac{\Gamma_{F} (P+Q)}{P^2 - m_e^2} \right] \cdot \nonumber \\
&  & \cdot \left( (1 - 4 sin^2 \theta_W) T_{\alpha \mu}^{S} \, + \,
i T_{\alpha \mu}^{A} \right) \, A^{\alpha} \gamma^{\mu}
\label{2.19}
\eea
where the tensors $T^{S,A}_{\alpha \mu}$ are defined in the appendix.
Again, from Lorentz invariance, we have that the
extra-contribution to the neutrino self-energy can be written as
\be
- \cal{R}e \widetilde{\Sigma^B} \; = \; a'_L \not{K} \, + \, b'_L \not{u}
\, + \, c_L \not{B}
\label{2.21b}
\ee
where $B_{\mu} \, = \, \frac{1}{2} \epsilon_{\mu \nu \alpha \beta} u^{\nu}
F^{\alpha \beta} \,= \, - i \epsilon_{\mu \nu \alpha \beta} u^{\nu}
Q^{\alpha} A^{\beta}$ and, in the rest frame of the medium,
$B_{\mu} \, = \, (0, \vec{B})$. To lowest order in
$G_F$, the explicit $K$-dependence in (\ref{2.18}), (\ref{2.19}) again
vanishes, and then $a'_L \, \simeq  \, 0$. For the coefficients
$b'_L$ and $c_L$, in the appendix it is shown that,
for simmetry reasons, also $b'_L \, \simeq \, 0$;
the charged current contribution to $c_L$, instead, can be written as
\be
c_L^W \; = \; - \frac{4 e G_F}{\sqrt{2}} \, \int \frac{d^3 p}{(2
\pi)^3} \, \frac{1}{2 E} \, \frac{d}{d E} \left( n_e \, - \, n_{\bar
e} \right)
\label{2.24}
\ee
where $E \, = \, \sqrt{p^2 \, + \, m_e^2}$ is the electron energy. The
neutral current contribution of the electron background can be
obtained in a similar way (see appendix):
\be
c_L^Z \; = \;  \frac{2 e G_F}{\sqrt{2}} \, \int \frac{d^3 p}{(2
\pi)^3} \, \frac{1}{2 E} \, \frac{d}{d E} \left( n_e \, - \, n_{\bar
e} \right)
\label{2.25}
\ee
Because of double sign change due to couplings with $Z^0$ and the
magnetic field of electrons and protons, the neutral current
contribution of the proton background is easily obtained by
(\ref{2.25}):
\be
c_L^Z \; = \;  \frac{2 e G_F}{\sqrt{2}} \, \int \frac{d^3 p}{(2
\pi)^3} \, \frac{1}{2 E} \, \frac{d}{d E} \left( n_p \, - \, n_{\bar
p} \right)
\label{2.26}
\ee
where now $E \, = \, \sqrt{p^2 \, + \, m_p^2}$ is the proton energy, and
$n_p$ ($n_{\bar p}$) is the Fermi distribution function for protons
(antiprotons).

Now, let us evaluate the integral
\be
I_e \; = \; \int \frac{d^3 p}{(2
\pi)^3} \, \frac{1}{2 E} \, \frac{d}{d E} \, n_e (E)
\label{2.27}
\ee
(the extension to protons is straigthforward) in the two physical
approximation outlined in the previous section.

Let us first consider the case of a degenerate electron gas ($T \ll
T_F$); for this
\be
\frac{d}{d E} \, n_e (E) \; \simeq \; - \, \delta (E - E_F)
\label{2.28}
\ee
where
\be
E_F \; = \; m_e \; + \; \frac{p_F^2}{2m_e} \; = \; m_e \; + \;
\frac{(3 \pi^2 N_e)^{\frac{1}{3}}}{2 m_e}
\ee
is the Fermi Energy, so that
\be
I_e \; \simeq \; - \, \frac{m_e}{4 \pi^2} \, \int_{m_e}^{+ \infty} \,
\frac{p(E)}{E} \, \delta (E-E_F) \, d E \; = \; - \, \frac{m_e}{4
\pi^2} \, \frac{p_F}{E_F}
\label{2.29}
\ee
and then
\be
I_e \; \simeq \; - \, \frac{p_F}{4 \pi^2}
\label{2.30}
\ee
For a neutral medium, the Fermi momentum of protons is equal to that
of electrons so that $I_p \, = \, I_e$.

Let us now consider a classical plasma, for which
\be
\frac{d}{d E} \, n_e (E) \; \simeq \; - \beta \, n_e (E)
\label{2.31}
\ee
In the non-relativistic approximation for the electrons in the
background,
\be
\frac{1}{E} \; \simeq \; \frac{1}{m_e} \, \left( 2 \, - \,
\frac{E}{m_e} \right)
\label{2.32}
\ee
we have
\be
I_e \; \simeq \; - \, \frac{\beta}{2 m_e} \, \left( N_e \, - \,
\frac{\ov E}{m_e} \right)
\label{2.33}
\ee
Where $\ov E$ is the mean thermal electron energy, and is given by
\be
\ov{E} \; = \; N_e m_e \, + \,  \int \frac{d^3 p}{(2
\pi)^3} \, \frac{p^2}{2 m_e} \, n_e \; \simeq \; N_e \, \left( m_e \,
+ \, \frac{3}{4 \beta} \right)
\label{2.34}
\ee
{}From (\ref{2.33}) and (\ref{2.34}) we obtain, for the electron background,
\be
I_e \; \simeq \,  \frac{3 N_e}{8 m_e^2}
\label{2.35}
\ee
Instead, the proton contribution is negligible:
\be
I_p \; \simeq \,  \frac{3 N_p}{8 m_p^2} \; \ll \; I_e
\label{2.36}
\ee
Finally, for a neutral medium, in the degenerate gas approximation we have got
\be
c_L \; \simeq \; 0
\label{2.37}
\ee
for $\nu_e$ and
\be
c_L \; \simeq \; - \, \frac{e G_F}{\sqrt{2}} \, \frac{(3 \pi^2 N_e)^{\frac{1}
{3}}}{\pi^2}
\label{2.38}
\ee
for $\nu_{\mu}$,$\nu_{\tau}$, while in the approximation of a
classical non-relativistic plasma
\be
c_L \; \simeq \; - \, \frac{3e G_F}{4 \sqrt{2}} \, \frac{N_e}{m_e^2}
\label{2.39}
\ee
for $\nu_e$ and \be
c_L \; \simeq \; + \, \frac{3e G_F}{4 \sqrt{2}} \, \frac{N_e}{m_e^2}
\label{2.40}
\ee
for $\nu_{\mu}$,$\nu_{\tau}$.
We now give an estimate of the order of magnitude of the previous
results. For $N_e \, \sim \, 10^{23} \, cm^{-3}$ we have
$|c_L| \, \sim \, 10^{-22} \, eV^{-1}$ for $\nu_{\mu}$, $\nu_{\tau}$
in a degenerate gas, and $|c_L| \, \sim \, 10^{-27} \, eV^{-1}$ for a
classical plasma. Comparing the Wolfestein term
$|b_L| \, \sim \, 10^{-14} \, eV$ with the energies
\bea
|c_L B| \; & \sim & \; 10^{-14} \, \left( \frac{B}{10^{10} \, Gauss}
\right) \, eV \;\;\;\;\;\;\;\;\;\;  from \; eq. \, (\ref{2.38})
\label{2.41}  \\
|c_L B| \; & \sim & \; 10^{-19} \, \left( \frac{B}{10^{10} \, Gauss}
\right) \, eV \;\;\;\;\;\;\;\;\;\;  from \; eqs. \,
(\ref{2.39}),(\ref{2.40})
\label{2.42}
\eea
we deduce that these terms are relevant in most astrophysical
(supernov\ae, white dwarf, etc.) and cosmological enviroments, in
which the magnetic field strength can be as large as $10^{10 \div 14}
\, Gauss$.

\section{The dispersion relation.}

The Dirac equation (\ref{2.1}) has a non trivial solution only if the
determinant $D(\omega,\vec{k},\vec{B})$ of the 4x4 matrix
$\not{K} - m - \Sigma$ vanishes. The neutrino dispersion relation is
then given by
\be
D(\omega,\vec{k},\vec{B}) \; = \; 0
\label{3.1}
\ee
With $\Sigma$ calculated in the previous section, at first order in
$G_F$, the relation (\ref{3.1}) reads
\be
\omega^2 \, - \, k^2 \; = \; m^2 \, - \, b_L (\omega + k)
\, + \, c_L (\omega + k) \frac{\vec{k} \cdot \vec{B}}{k}
\label{3.2}
\ee
for the usual left-handed neutrinos, and
\be
\omega^2 \, - \, k^2 \; = \; m^2 \, - \, b_L (\omega - k)
\, - \, c_L (\omega - k) \frac{\vec{k} \cdot \vec{B}}{k}
\label{3.3}
\ee
for the right-handed neutrinos. Observe that the presence of a
privileged direction, that of the magnetic field $\vec{B}$, makes
non-isotropic the dispersion relations (\ref{3.2}),(\ref{3.3}):
besides a temporal dispersion in $\omega$ there is also a spatial
dispersion in $\vec{k}$ (also for isotropic media). Moreover, let us
note that, on the contrary to that happens in absence of a magnetic
field, making the transformation $k \rightarrow - k$ in
(\ref{3.2}),(\ref{3.3}) these two relations don't transform each
other; this happens only if one transforms $\vec{B}$ in $- \vec{B}$.
This point will be discussed in the next section.

For ultrarelativistic neutrinos, the left-handed ones propagate in
matter with energy given by
\be
\omega \; \simeq \; k \, + \, \frac{m^2}{2k} \, - \, b_L \, + \, c_L
\, \frac{\vec{k} \cdot \vec{B}}{k}
\label{3.4}
\ee
while right-handed neutrinos, not having coherent interactions with
matter, propagate as in vacuum:
\be
\omega \; \simeq \; k \, + \, \frac{m^2}{2k}
\label{3.5}
\ee
The action of the thermal background on neutrino propagation can be
described ``macroscopically" by an effective potential $V$ experienced
by neutrinos. This potential can be defined \cite{Nieves} as the
difference between the total energy and the kinetic energy of
neutrinos, that is
\be
V \; = \, - \, b_L \, + \, c_L
\, \frac{\vec{k} \cdot \vec{B}}{k}
\label{3.6}
\ee
For $\vec{B} = 0$ we obtain the Wolfestein result \cite{3}.

An interesting feature comes out for a particular configuration of the
magnetic field, given by the relation
\be
b_L \, k \; = \; c_L \, \vec{k} \cdot \vec{B}
\label{3.7}
\ee
As one can see from (\ref{3.4}) or (\ref{3.6}), if equation
(\ref{3.7}) holds, the coherent effects due properly to the magnetic
field are opposite to those generated in absence of the field, and
then the (left-handed) neutrinos propagate in matter as in vacuum.
Let us examine the conditions under which (\ref{3.7}) can be verified.

First of all, we observe that (\ref{3.7}) is almost independent on
neutrino energy; this dependence can occur only through $b_L$,$c_L$ at
order next to $G_F$. For propagation in a degenerate gas, from eqs.
(\ref{2.14a}),(\ref{2.14b}) and (\ref{2.37}),(\ref{2.38}) we deduce
that electron-neutrinos cannot experience the free propagation
condition (\ref{3.7}), while this condition can be verified by $\mu$-
and $\tau$-neutrinos only if $\vec{k} \cdot \vec{B}$ is negative.
Instead, all neutrino flavours can undergo the condition (\ref{3.7})
in a non-relativistic classical plasma for $\vec{k} \cdot \vec{B}$
positive. Note that the free propagation condition doesn't take place
for propagation normal to the magnetic field (in this case the
magnetic field has no effect on the dispersion relation).

For $\nu_{\mu}$,$\nu_{\tau}$ in a metal ($N_e \simeq 10^{23} \,
cm^{-3}$, $Z/A \simeq 1/2$) the field strength necessary to satisfy
(\ref{3.7}) scales as the power 2/3 of the density, and is
approximatively $B \sim 10^8 \, Gauss$. A very interesting thing
happens, instead, for neutrino propagation in a classical
non-
\newpage
\noindent
relativistic plasma: the field strength for (\ref{3.7}) doesn't
depend on medium characteristic \footnote{A very weak dependence on
medium is expressed in $m_e$, which would be the ``effective" electron
mass in matter.}
\be
B \; \simeq \; \frac{|b_L|}{|c_L|} \; \simeq \frac{4}{3} \,
\frac{m_e^2}{e}
\label{3.8}
\ee
Let us note that, apart a numerical factor, the field strength
(\ref{3.8}) corresponds to the Landau critical field
$B_c \, = \, m^2/e  \, \simeq \, 4.4 \cdot 10^{13} \, Gauss$
\cite{Landau} at which quantum effects become important in classical
electrodynamics \footnote{This limit can be obtained by equating
$\hbar \omega_B \, = \, m_e c^2$ where $\omega_B \, = \, \frac{eB}{m_e
c}$ is the gyrofrequency.}

\section{Spatial dispersion.}

In the previous section it has been pointed out that, also for isotropic
media, because of the presence of a magnetic field, the dispersion relation
(\ref{3.1}) doesn't manifest spatial isotropy, but
it depends on the relative direction of the neutrino momentum $\vec{k}$
and the magnetic field $\vec{B}$ (spatial dispersion). This means
that the eigen-energies given by (\ref{3.2}),(\ref{3.3}) don't
specify completely the eigen-modes of propagation of neutrinos in
magnetized media: there are also eigen-directions to be picked out.
To find these, we can follow two different ways.

A direct method consists in the application of the Ward identity at
finite temperature \cite{WT} (this can be applied only for static
fields). Let us rewrite the temperature-dependent real parts of the
graphs in fig. 1,3 as
\bea
\cal{R}e \widetilde{\Sigma_W} \; & = & \; \frac{g^2}{2} \,
\int \frac{d^4 P}{(2 \pi)^4} \,
\frac{1}{P^2 - M_W^2} \,
\gamma_{\alpha} \,  \left( \not{P}
\, + \, \not{K} \, + \, m_e \right) \, \gamma^{\alpha} \cdot \nonumber \\
&  & \cdot 2 \pi \, n_F(P + K) \, \delta \left( (P
+ K)^2 \, - \, m_e^2 \right)
\label{4.1}  \\
\vspace{1cm}
\cal{R}e \widetilde{\Sigma_W^B} \; & = & \; \frac{e \, g^2}{2} \,
\int \frac{d^4 P}{(2 \pi)^4} \,
\frac{1}{P^2 - M_W^2} \,
\gamma_{\alpha} \, \left( \not{P} \, + \, \not{K} \, + \, m_e \right)
\gamma_{\mu} \, \left( \not{P} \, + \, \not{K} \, + \, m_e \right)
\gamma^{\alpha} \cdot \nonumber \\
&  & \cdot 4 \pi \, n_F(P + K) \,
\frac{\delta ( (P + K)^2 \, - \, m_e^2)}{(P + K)^2 \,
- \, m_e^2} \, A^{\mu}
\label{4.2}
\eea
(for simplicity we consider a medium without antiparticles). Deriving
(\ref{4.1}) with respect to $K_{\mu}$,
and substituting in this expression the relation \cite{WT}
\newpage
\bea
- \, \frac{\partial}{\partial P_{\mu}}
\, \left( 2 \pi n_F(P)
\, \delta (P^2 \, - \, m_e^2) \right) & = &
\left( \not{P} \, + \, m_e \right) \, \gamma_{\mu}
\left( \not{P} \, + \, m_e \right) \,
\left[ 4 \pi n_F(P) \frac{\delta (P^2 \, - \, m_e^2 )}
{P^2 \, - \, m_e^2} \, \right] \; + \;  \nonumber \\
&  & - \, 2 \pi \frac{\partial n_F(P)}{\partial P_{\mu}}
\left( \not{P} \, + \, m_e \right) \, \delta(
P^2 \, - \, m_e^2) \;\; ,
\label{4.3}
\eea
for $A_{\mu} \, = \, (0, \vec{A})$, from (\ref{4.2}) we obtain the
following identity
\be
\cal{R}e \widetilde{\Sigma_W^B} \; = \; - e \, A_{\mu} \,
\frac{\partial}{\partial K_{\mu}} \cal{R}e \widetilde{\Sigma_W} (K)
\label{4.4}
\ee
This relation is very similar to the usual Ward identity at $T=0$ in
Q.E.D. \cite{Ward}; however, its applicability domain is much smaller
than that at zero temperature (for a detailed discussion on the
Ward identity
at finite temperature see \cite{WT} and references therein).
At first order in $G_F$, from (\ref{2.7}) and (\ref{2.21b}) we deduce
\be
\left( 2e (K \cdot A)
\, \left( \frac{\partial a_L}{\partial y} \, K_{\mu} \, + \,
\frac{\partial b_L}{\partial y} \, u_{\mu} \right)
\; + \; c_L \, B_{\mu}
\right) \, \gamma^{\mu} \; \simeq \; 0
\label{4.5}
\ee
{}From the independence of the Dirac matrices, we finally arrive to the
following approximate equalities:
\be
\frac{\partial b_L}{\partial y} \; + \; \omega \,
\frac{\partial a_L}{\partial y} \; \simeq \; 0
\label{4.6}
\ee

\be
2e (K \cdot A) \, \frac{\partial a_L}{\partial y} \, \vec{k}
\; \simeq \; c_L \, \vec{B}
\label{4.7}
\ee
The relation (\ref{4.6}) is nothing that (\ref{2.11}) at lowest order in
$G_F$ (at this order a straigthforward calculation gives
$\frac{\partial T_u}{\partial y} \, \simeq \, 0$). From (\ref{4.7}),
after simple manipulations, we can write \footnote{At order $G_F$ a
direct computation gives $\frac{\partial a_L}{\partial y} \, \simeq \,
\frac{G_F}{\sqrt{2}} \, \frac{\omega}{k^4} \, N_e$.}
\be
\frac{\vec{k}}{k} \; = \; \pm \, \frac{\vec{B}}{B}
\label{4.8}
\ee
Then, from (\ref{4.8}), we deduce that the eigen-directions of
neutrinos in the medium are the directions parallel and antiparallel
to the applied magnetic field (the $+$ sign refers to $\nu_L$ while
$-$ to $\nu_R$).

The physical interpretation of this result comes again from the Dirac
equation, that in the Weyl representation reads
\bea
\left( \omega \; + \; b_L \; + \vec{\sigma} \cdot \left( \vec{k} \; +
\; c_L \, \vec{B} \right) \right) \; \psi_L \; & = & \; m \, \psi_R
\label{4.9} \\
\vspace{1cm}
\left( \omega \; - \; \vec{\sigma} \cdot  \vec{k} \right) \;
\psi_R \; & = & \; m \, \psi_L
\label{4.10}
\eea
By defining the helicity eigen-states
\be
\frac{\vec{\sigma} \cdot  \vec{k}}{| \vec{k} |} \; \phi_{\lambda} \; =
\; \lambda \, \phi_{\lambda}
\label{4.11}
\ee
with $\lambda \, = \, \pm 1$, because the interaction here considered
through the diagrams in fig. 3,4 doesn't induce helicity flip
\footnote{Terms in the self-energy that change helicity are of the
form $d_L \, \frac{1 \, - \, \gamma_5}{2} \, + \, d_R \, \frac{1 \,
+ \, \gamma_5}{2}$, but in the Standard Model at order one-loop the
coefficients $d_L$,$d_R$ are vanishing (here we not consider intrinsic
neutrino magnetic moment). In scenarios beyond the Standard Model,
with more Higgs particles, these coefficients can be non zero also at
order one-loop (see for example \cite{Pakist}).},
the spinors $\psi_L$,$\psi_R$ must be proportional to
$\phi_{\lambda}$. From eq. (\ref{4.9}) we deduce that this evenience
can take place only if the operators $\vec{\sigma} \cdot  \vec{k}$ and
$\vec{\sigma} \cdot  \vec{B}$ commute between them, that is
(\ref{4.8}) holds.

In other words, the neutrino eigen-directions, parallel to the
magnetic field, are a direct consequence of helicity conservation of
the interaction.

A peculiar feature of the spatial dispersion is the deflection of
neutrino momentum at the exit of the medium. For evaluating this, let
us consider the neutrino velocity $\vec{v}$, whose components are
given by
\be
v_i \; = \; \frac{\partial \omega}{\partial k_i}
\label{4.12}
\ee
{}From eq. (\ref{3.2}) we then obtain
\be
\vec{v} \; = \; \frac{1}{\omega} \, \left( \vec{k} \, + \, \vec{k'}
\right)
\label{4.13}
\ee
with
\be
\vec{k'} \; = \; - \, b_L \, \frac{\vec{k}}{k} \; + \; c_L \, \vec{B}
\label{4.14}
\ee
For $\vec{B} \, = \, 0$ we recover the fact that there is no spatial
dispersion (only the modulus of the velocity changes), while for
$\vec{B}$ and $\vec{k}$ parallel between them there is no deflection,
so that we reobtain the eigen-directions (\ref{4.8}).

In general, for $\vec{k} \times \vec{B} \, \neq \, 0$ neutrinos are
deflected from initial trajectory; if, for simplicity, we consider
$\vec{k}$ along the z-axis ($\vec{k} = (0,0,k)$) and $\vec{B}$
in the xz-plane ($\vec{B} = (B_x,0,B_z)$) the angle $\theta$ between
the final and the initial direction is given by
\be
tg \, \theta \; = \; \frac{v_x}{v_z} \; = \; \frac{c_L \, B_x}{k} \,
\frac{1}{1 \, - \, \frac{b_L}{k} \, + \, \frac{c_L \, B_z}{k}} \;
\simeq \; \frac{c_L \, B_x}{k} \; \ll \; 1
\label{4.15}
\ee
In practice, for all physical situations this deflection is enormously
small to detect.

\section{Effects on neutrino matter oscillations.}

In general if neutrinos are massive, the flavour (vacuum) eigen-states
$\nu_{\alpha}$ ($\alpha \, = \, e, \mu , \tau $) can be expressed as a
linear superposition of the mass eigen-states $\nu_i$ ($i \, = \,
1,2,3$): $\nu_{\alpha} \, = \, \sum_i \, U_{\alpha i} \, \nu_i$. Let
us consider, for simplicity, the mixing of only two flavours ($e$ and
$\alpha \, = \, \mu$ or $\tau$); in this case the mixing matrix $U$,
if $CP$ is conserved, can be cast in the form
\be
U \; = \; \left( \ba{cc}
                 cos \, \theta    &  - \, sin \, \theta  \\
                 sin \, \theta    &  cos \, \theta
                 \ea  \right)
\label{5.1}
\ee
where $\theta$ is the vacuum mixing angle. The hamiltonian of the
sistem of two neutrinos that has to be considered for the propagation
of a mass eigen-state in a magnetized medium has, in the
ultrarelativistic approximation, the form
\be
H \; = \; k \; + \; \frac{m^2}{2k} \; - \; b_L \; + \; c_L \,
\frac{\vec{k} \cdot \vec{B}}{k}
\label{5.2}
\ee
where, in the mass eigen-states basis,
\be
\frac{m^2}{2k} \; = \; \left(  \ba{cc}
                               \frac{m_1^2}{2k}  &  0 \\
                               0  &  \frac{m_2^2}{2k}
                               \ea  \right)
\label{5.3}
\ee

\be
b_L \; = \; b_L^W \, \left( \ba{cc}
                 cos^2 \, \theta & - \, sin \, \theta  \, cos \, \theta \\
                 - \, sin \, \theta  \, cos \, \theta  &  sin^2 \, \theta
                 \ea  \right)
 \; + \; b_L^Z \, \left( \ba{cc}
                         1  &  0  \\
                         0  &  1
                         \ea  \right)
\label{5.4}
\ee

\be
c_L \; = \; c_L^W \, \left( \ba{cc}
                 cos^2 \, \theta & - \, sin \, \theta  \, cos \, \theta \\
                 - \, sin \, \theta  \, cos \, \theta  &  sin^2 \, \theta
                 \ea  \right)
 \; + \; c_L^Z \, \left( \ba{cc}
                         1  &  0  \\
                         0  &  1
                         \ea  \right)
\label{5.4}
\ee
and $b_L^W$,$b_L^Z$ are taken respectively by (\ref{2.13}),(\ref{2.14})
while $c_L^W$ by (\ref{2.24}) and $c_L^Z$ by the sum of (\ref{2.25})
and (\ref{2.26}).

The eigen-energies of propagating neutrinos are given, then, by the
eigen-values of $H$:
\bea
\omega_{1,2} \; & = & \; k \; + \; \frac{1}{2} \, \frac{m^2_1 \, + \,
m^2_2}{2k} \; - \; \frac{1}{2} \, \left( b_L^W \, - \, c_L^W
\frac{\vec{k} \cdot \vec{B}}{k} \right) \; - \; \left( b_L^Z \, - \, c_L^Z
\frac{\vec{k} \cdot \vec{B}}{k} \right) \; +  \nonumber  \\
&  & \overline{+} \; \frac{1}{2} \, \sqrt{\left(
\frac{\Delta m^2}{2k} \, cos \, 2
\theta \; + \; b_L^W \, - \, c_L^W \frac{\vec{k} \cdot \vec{B}}{k}
\right)^2 \; + \; \left( \frac{\Delta m^2}{2k} \, sin \, 2
\theta \right)^2}
\label{5.6}
\eea
where $\Delta m^2 \, = \, m_2^2 \, - m_1^2$. For $\theta \, = \, 0$ we
recover (\ref{3.4}).

The matter eigen-states $\nu_{1m}$,$\nu_{2m}$ again can be expressed
as a linear superposition of the flavour eigen-states, with an
effective mixing angle $\theta_m$ given by
\be
sin \, 2 \theta_m \; = \; \frac{ \frac{\Delta m^2}{2k} \, sin \, 2
\theta}{\sqrt{\left( \frac{\Delta m^2}{2k} \, cos \, 2
\theta \; + \; b_L^W \, - \, c_L^W \frac{\vec{k} \cdot \vec{B}}{k}
\right)^2 \; + \; \left( \frac{\Delta m^2}{2k} \, sin \, 2
\theta \right)^2}}
\label{5.7}
\ee
Because of flavour superposition, if for example one creates a $\nu_e$
by weak interactions and then sends it in a magnetized medium, at the
exit of the medium there is a non-vanishing probability of finding a
neutrino of another flavour. For a costant density medium, the transition
probability at a certain distance $x$ from the source is given by
\footnote{The calculation of the transition probability proceeds as in
the MSW theory without a magnetic field, but with an effective mixing
angle given by (\ref{5.7}). See for example \cite{Prob}.}
\be
P(\nu_e \rightarrow \nu_{\mu}) \; = \; sin^2 \, \theta_m  \; sin^2 \,
\frac{\pi x}{L_m}
\label{5.8}
\ee
where $L_m$ is the effective oscillation length in matter,
\be
L_m \; = \; \frac{2 \pi}{\omega_2 \, - \, \omega_1} \; = \; \frac{2
\pi}{\sqrt{\left( \frac{\Delta m^2}{2k} \, cos \, 2
\theta \; + \; b_L^W \, - \, c_L^W \frac{\vec{k} \cdot \vec{B}}{k}
\right)^2 \; + \; \left( \frac{\Delta m^2}{2k} \, sin \, 2
\theta \right)^2}}
\label{5.9}
\ee
{}From (\ref{5.7}) we observe that the MSW resonance condition is
changed: in presence of a magnetic field the maximum effect of matter
on neutrino oscillation (maximum effective mixing angle) occurs if the
relation
\be
\frac{\Delta m^2}{2k} \, cos \, 2
\theta \; = \; - \, b_L^W \; + \; c_L^W \frac{\vec{k} \cdot \vec{B}}{k}
\label{5.10}
\ee
holds. More explicitely, for a degenerate gas the resonance condition
is
\be
\frac{\Delta m^2}{2k} \, cos \, 2 \theta \; = \;
\sqrt{2} \, G_F \, N_e \; + \; \frac{e G_F}{\sqrt{2}} \,
\frac{(3 \pi^2 N_e)^{\frac{1}{3}}}{\pi^2} \,
\frac{\vec{k} \cdot \vec{B}}{k}
\label{5.11}
\ee
while for a classical non-relativistic plasma we have
\be
\frac{\Delta m^2}{2k} \, cos \, 2 \theta \; = \;
\sqrt{2} \, G_F \, N_e \; - \; \frac{3e G_F}{2 \sqrt{2}} \, \frac{N_e}{m_e^2}
\, \frac{\vec{k} \cdot \vec{B}}{k}
\label{5.12}
\ee
The effect of the magnetic field, as it has already been pointed out,
vanishes for propagation normal to the field and, in the other cases,
is important for strong fields.

A novel feature appear when the magnetic field configuration is such
that
\be
b_L^W \, k \; = \; c_L^W \, \vec{k} \cdot \vec{B}
\label{5.13}
\ee
In this case, neutrinos don't propagate freely in matter
\footnote{From (\ref{5.2}), for free propagation in matter, not only
(\ref{5.13}) but also the relation
\be
\frac{b_L^W}{b_L^Z} \; = \; \frac{c_L^W}{c_L^Z}
\label{5.14}
\ee
between the coefficients must be realized.}
but no resonance occurs. In fact, if (\ref{5.13}) holds, the
hamiltonian (\ref{5.2}) is diagonal, so that no effective mixing angle
$\theta_m$ arises. The conditions under which (\ref{5.13}) can be
verified are similar to those obtained in section 3.

\section{Discussions and conclusions.}

In the present work we have studied the modification of the neutrino
``effective properties'' in an isotropic medium when a magnetic field
is present. Because of the privileged direction introduced by the
magnetic field $\vec{B}$, spatial dispersion arises so that for
individuating the eigen-modes of neutrino propagation one has to
recognize also the eigen-directions of neutrinos. From the helicity
conservation in the interaction between neutrinos and the magnetized
matter, we have that the eigen-directions are those parallel to the
magnetic field. A direct consequence of this spatial dispersion is the
deflection of the neutrino momentum at the exit of the medium;
however, a simple calculation (see (\ref{4.15})) shows that the
deflection is practically no detectable in all physical situations.

An important result is the modification of the effective potential
(\ref{3.6}) experienced by neutrinos in matter under the influence of
the magnetic field. For neutrino matter oscillations, this implies
that the MSW resonance condition changes according to (\ref{5.10})
for sufficiently high magnetic fields. This can be easily visualized
with the level crossing diagrams showed in figs. 5-7. The resonance
point change with the field strength and with the field direction,
reaching maximum modification for neutrino propagation along the
magnetic field, while no modification occurs for normal propagation.
In particular, if the field configuration is such that the charged
current contribution to neutrino effective potential cancel
themselves, no difference arises in the propagation of the two
neutrino eigen-states, so that no level crossing occurs and then no
resonance can be present. Moreover, for this field configuration, if
we consider a (neutral) degenerate gas with
\be
N_e \; - \; N_{\bar e} \; = \; \frac{1}{2} \, \left(
N_n \; - \; N_{\bar n} \right)
\label{6.1}
\ee
or a classical non-relativistic plasma with
\be
N_e \; - \; N_{\bar e} \; = \; \left(
N_n \; - \; N_{\bar n} \right)
\label{6.2}
\ee
the medium is completely transparent to neutrinos (see (\ref{5.14})).
Let us note that the no resonance condition (\ref{5.13}) and the
additional relation (\ref{5.14}) for free propagation are
independent on neutrino energy. Furthermore, for those physical
situations in which the classical plasma approximation is valid, the
strength of the magnetic field necessary to satisfy (\ref{5.13})
doesn't depend on medium characteristic but, apart a numerical factor,
coincides with the Landau critical field $B_c \, = \, \frac{m_e^2}{e}
\, \simeq \, 4.4 \cdot 10^{13} \, Gauss$.

These considerations should be kept in mind when one studies neutrino
propagation in astrophysical and cosmological enviroments. In fact,
for example, even if one ignores specific details, a magnetic field is
present in supernov\ae~ with strengths up to $10^{14} \, Gauss$
\cite{supernovae}: this implies that resonant MSW oscillations of
neutrinos through their interactions with background electrons,
protons and neutrons (but not with background neutrinos) can no longer
occur. Moreover, the possibility of completely free propagation in
these backgrounds can alter significatively the dynamics of the
supernov\ae.

Instead, MSW oscillations of solar neutrinos \cite{4} are practically
no interested by our results because of too small magnetic fields
present in the Sun \cite{sun}.

Another physical context in which one can use our results is the
evolution of the Universe. In fact, a magnetic field close to the
critical field $B_c$ is present \cite{univ} just before the epoch of
the nucleosynthesis of the light elements, and the number of each type
of neutrinos is important for the development of the nucleosynthesis
itself \cite{nucleos}. \\

\newpage

\noindent
{\Large \bf Appendix}\\

\begin{appendix}
\section{Calculation details.}

Let us evaluate the neutrino self-energy in a magnetized medium by
starting with charged-current diagram in fig. 3. We follow the lines
traced by D'Olivo, Nives and Pal in Ref. \cite{3}.

At lowest order in $G_F$, the temperature-dependent real part of the
self-energy is given by:
\bea
\cal{R}e \Sigma^B_W \; & = & \; - \, \lim_{|\vec{q}| \rightarrow 0}
\frac{4 e G_F}{\sqrt{2}} \, \int \frac{d^4 P}{(2 \pi)^4} \, \gamma_{\mu}
\, \frac{1 - \gamma_5}{2} \, \left( \not{P} \, + \, \not{Q} \,+ \, m_e
\right) \, \gamma_{\alpha} \, \left( \not{P} \, + \, m_e \right) \,
\gamma^{\mu} \, \frac{1 - \gamma_5}{2} \cdot   \nonumber \\
&  & \cdot \left[ \frac{\Gamma_F (P)}{\left( P \, + \, Q \right)^2 \, - \,
m_e^2} \; + \; \frac{\Gamma_F (P \, + \, Q)}{P^2 \, - \, m_e^2}
\right] \; A^{\alpha}
\label{a.1}
\eea
By writing
\be
\gamma_{\mu}
\, \frac{1 - \gamma_5}{2} \, \left( \not{P} \, + \, \not{Q} \,+ \, m_e
\right) \, \gamma_{\alpha} \, \left( \not{P} \, + \, m_e \right) \,
\gamma^{\mu} \, \frac{1 - \gamma_5}{2} \; = \; - \, T_{\alpha \mu} \,
\gamma^{\mu} \, \frac{1 - \gamma_5}{2}
\label{a.2}
\ee
with
\be
T_{\alpha \mu} \; \equiv \; Tr \left[
\left( \not{P} \, + \, \not{Q} \,+ \, m_e
\right) \, \gamma_{\alpha} \, \left( \not{P} \, + \, m_e \right) \,
\gamma^{\mu} \, \frac{1 - \gamma_5}{2} \right] \; = \;
2 \left( T^S_{\alpha \mu} \; + \; i \, T^A_{\alpha \mu} \right)
\label{a.3}
\ee

\be
T^{S}_{\alpha \mu} \; = \; 2 P_{\alpha} P_{\mu} \, + \, Q_{\alpha}
P_{\mu} \, + \, P_{\alpha} Q_{\mu} \, - \, (P^2 - m_e^2) \, g_{\alpha
\mu} \, - \, P \cdot Q \, g_{\alpha \mu}
\label{a.4}
\ee

\be
T_{\alpha \mu}^{A} \; = \; \epsilon_{\alpha \mu \delta \beta} \,
P^{\delta} Q^{\beta}
\label{a.5}
\ee
one obtains:
\bea
\cal{R}e \widetilde{\Sigma^B_W} & = & \lim_{|\vec{q}| \rightarrow 0}
\frac{8 e G_F}{\sqrt{2}} \, \int \frac{d^4 P}{(2 \pi)^4} \,
2 \pi \, \delta (P^2 \, - \, m_e^2) \left[
\theta (P^0) \, n_e(P) \; + \; \theta (- \, P^0) \,
n_{\bar e}(P) \right] \cdot  \nonumber \\
&  & \cdot \left[ \left( \frac{T^S_{\alpha \mu} (Q)}
{P^2 \, - \, m_e^2 \, +
\, Q^2 \, + \, 2 P \cdot Q} \; + \; \left( Q \; \rightarrow \; - \, Q
\right) \right) \; + \right. \nonumber  \\
&  & \left. + \; i \, T^A_{\alpha \mu} (Q) \, \left( \frac{1}{
P^2 \, - \, m_e^2 \, +
\, Q^2 \, + \, 2 P \cdot Q} \; + \; \left( Q \; \rightarrow \; - \, Q
\right) \right) \right] \, A^{\alpha} \, \gamma^{\mu}
\label{a.6}
\eea
Eliminating the $\delta$-function with the energy integration, we have
\bea
\cal{R}e \widetilde{\Sigma^B_W} & = & \lim_{|\vec{q}| \rightarrow 0}
\frac{8 e G_F}{\sqrt{2}} \, \int \frac{d^3 p}{(2 \pi)^3} \,
\frac{1}{2E} \, \cdot \nonumber \\
&  & \cdot \left[ \left(n_e \, + \, n_{\bar e} \right) \,
\left( \frac{2 P_{\alpha} P_{\mu} \, + \, Q_{\alpha}
P_{\mu} \, + \, P_{\alpha} Q_{\mu} \,
- \, P \cdot Q \, g_{\alpha \mu}}{Q^2 \, + \, 2 P \cdot Q}
\; + \; \left( Q \rightarrow - Q \right) \right) \; +
\right. \nonumber \\
&  & \left. + \;  \left(n_e \, - \, n_{\bar e} \right) \, i \,
\epsilon_{\alpha \mu \delta \beta} \, P^{\delta} Q^{\beta} \,
\left( \frac{1}{Q^2 \, + \, 2 P \cdot Q}
\; + \; \left( Q \; \rightarrow \; - \, Q \right) \right) \right]
\, A^{\alpha} \, \gamma^{\mu}
\label{a.7}
\eea
The dispersion relation must be electromagnetic gauge-invariant; to
ensure this, let us introduce the 4-vector
$B_{\mu} \, = \, \frac{1}{2} \epsilon_{\mu \nu \alpha \beta} u^{\nu}
F^{\alpha \beta}$
that in the rest frame of the medium has component $B_{\mu} \, = \,
(0, \vec{B})$. From Lorentz invariance, because (\ref{a.7}) depends
only on the 4-vectors $u_{\mu}$, $B_{\mu}$, one can write
\be
- \cal{R}e \widetilde{\Sigma^B} \; = \; b'_L \not{u}
\, + \, c_L \not{B}
\label{a.8}
\ee
where, from $u \cdot B \, = \, 0$, the coefficients are given by
\be
b'_L \; = \; - \, \frac{1}{4} \, Tr(\not{u} \cal{R}e
\widetilde{\Sigma^B})
\label{a.9}
\ee

\be
c_L \; = \; - \, \frac{1}{4 B^2} \, Tr(\not{B} \cal{R}e
\widetilde{\Sigma^B})
\label{a.10}
\ee
{}From gauge-invariance we have that $Q \cdot A \, = \, 0$ and, for a
static field, $Q \cdot u \, = \, 0$. Moreover, by choosing $A_{\mu} \,
= \, (0, \vec{A})$, also $u \cdot A \, = \, 0$ holds. From these, the
coefficient $b'_L$ can be written as
\bea
b'_L \; & = &\; \lim_{|\vec{q}| \rightarrow 0}
\frac{8 e G_F}{\sqrt{2}} \, \int \frac{d^3 p}{(2 \pi)^3} \,
\left[ \left(n_e \, + \, n_{\bar e} \right) \,
\vec{p} \cdot \vec{A} \; + \right.\nonumber \\
&  & \left. + \; \left(n_e \, - \, n_{\bar e} \right) \,
\frac{\vec{p} \cdot \vec{B}}{2E} \, \left( \frac{1}{2 \vec{p}
\cdot \vec{q} \, - \, \vec{q}^2} \; - \;  \frac{1}{2 \vec{p}
\cdot \vec{q} \, + \, \vec{q}^2}
\right) \right]
\label{a.11}
\eea
Let us first perform the integration on the azimuthal $\theta$-angle;
because the integrand function is odd under the exchange $cos \,
\theta \, \rightarrow - \, cos \, \theta$, the integration yields
\be
b'_L \; = \; 0
\label{a.12}
\ee
For the coefficient $c_L$ one instead gets:
\bea
c_L & = & \lim_{|\vec{q}| \rightarrow 0}
\frac{8 e G_F}{\sqrt{2}} \, \left[ \int \frac{d^3 p}{(2 \pi)^3} \,
\left(n_e \, + \, n_{\bar e} \right) \,
\frac{(\vec{p} \cdot \vec{A}) \,(\vec{p} \cdot \vec{B})}{\vec{B}^2}
\frac{1}{E} \, \left( \frac{1}{2 \vec{p}
\cdot \vec{q} \, - \, \vec{q}^2} \; - \;  \frac{1}{2 \vec{p}
\cdot \vec{q} \, + \, \vec{q}^2} \right) \; + \right.\nonumber \\
&  & \left. - \;\frac{1}{2} \, \int \frac{d^3 p}{(2 \pi)^3} \,
\left(n_e \, - \, n_{\bar e} \right) \,
\left( \frac{1}{2 \vec{p}
\cdot \vec{q} \, - \, \vec{q}^2} \; - \;  \frac{1}{2 \vec{p}
\cdot \vec{q} \, + \, \vec{q}^2}
\right) \right]
\label{a.13}
\eea
In the first integral, the integrand function is again odd in $cos \,
\theta$, so that one remains with
\be
c_L \; = \; - \, \lim_{|\vec{q}| \rightarrow 0}
\frac{4 e G_F}{\sqrt{2}} \, \int \frac{d^3 p}{(2 \pi)^3} \,
\left(n_e \, - \, n_{\bar e} \right) \,
\left( \frac{1}{2 \vec{p}
\cdot \vec{q} \, - \, \vec{q}^2} \; - \;  \frac{1}{2 \vec{p}
\cdot \vec{q} \, + \, \vec{q}^2} \right)
\label{a.14}
\ee
Because the function in the integral of eq. (\ref{a.14}) is uniformly
sommable, we can perform first the limit $\vec{q} \rightarrow 0$
(fixed the direction of $\vec{q}$); for this purpose let us make the
substitutions $\vec{p} \, \rightarrow \, \vec{p} \, + \, \frac{1}{2}
\, \vec{q}$ in the first term of (\ref{a.14}) and
$\vec{p} \, \rightarrow \, \vec{p} \, - \, \frac{1}{2}
\, \vec{q}$ in the second one:
\be
c_L \; = \; - \, \frac{4 e G_F}{\sqrt{2}} \,
\int \frac{d^3 p}{(2 \pi)^3} \, \lim_{|\vec{q}| \rightarrow 0}
\, \frac{1}{2 \vec{p} \cdot \vec{q}} \, \left( \left[
n_e(E_+) \, - \, n_{e}(E_-) \right] \; - \; \left[
n_{\bar e}(E_+) \, - \, n_{\bar e}(E_-) \right] \right)
\label{a.15}
\ee
Neglecting terms of the second order in $|\vec{q}|$,
\be
E_{\pm} \; \equiv \; \sqrt{\left( \vec{p} \, \pm \, \frac{1}{2}
\, \vec{q} \right)^2 \; + \; m_e^2} \; \simeq \; E \; \pm \;
\frac{1}{2} \, \frac{\vec{p} \cdot \vec{q}}{E}
\label{a.16}
\ee
one obtains
\bea
c_L \; = \; - \, \frac{4 e G_F}{\sqrt{2}} \,
\int \frac{d^3 p}{(2 \pi)^3} \,
& \left[ \lim_{h \rightarrow 0}
\frac{n_e (E \, + \, \frac{1}{2} \, h) \; - \;
n_e(E \, - \, \frac{1}{2} \, h)}{h} \; + \right. \\
& \left. - \; \lim_{h \rightarrow 0}
\frac{n_{\bar e} \left( E \, + \, \frac{1}{2} \, h \right)
 \; - \; n_{\bar e} \left( E \, - \, \frac{1}{2} \, h \right)}{h} \right]
\label{a.17}
\eea
Performing the limits, we finally arrive at
\be
c_L \; = \; - \, \frac{4 e G_F}{\sqrt{2}} \,
\int \frac{d^3 p}{(2 \pi)^3} \,
\frac{1}{2E} \, \frac{d}{dE} \,
\left( n_e \, - \, n_{\bar e} \right)
\label{a.18}
\ee
that is eq.(\ref{2.24}).

For the neutral-current contribution (for example from electron
background), from the identity (\ref{a.2}) we have
\be
\cal{R}e \widetilde{\Sigma^B_Z} \; = \; - \,
\lim_{|\vec{q}| \rightarrow 0}
\frac{2 e G_F}{\sqrt{2}} \, \int \frac{d^4 P}{(2 \pi)^4} \,
\left[ \frac{\Gamma_F (P)}{\left( P \, + \, Q \right)^2 \, - \,
m_e^2} \; + \; \frac{\Gamma_F (P \, + \, Q)}{P^2 \, - \, m_e^2}
\right] \;
T'_{\alpha \mu} \, A^{\alpha} \, \gamma^{\mu}
\label{a.19}
\ee
where
\be
T'_{\alpha \mu} \; = \;
2 \left[ \left( 1 \, - \, 4 \, sin^2 \, \theta_W \right) \,
T^S_{\alpha \mu} \; + \; i \, T^A_{\alpha \mu} \right]
\label{a.20}
\ee
and $T^{S,A}_{\alpha \mu}$ are given by (\ref{a.4}),(\ref{a.5}). From
these, it is simple to observe that also the neutral-current
contribution to $b'_L$ vanishes while, because the simmetric part of
$T'_{\alpha \mu}$ always vanishes in the integration over $cos \,
\theta$, the contribution to $c_L$ can be obtained by the
charged-current one by substituting $G_F$ with $- \, G_F/2$.
\end{appendix}

\vspace{1truecm}

\noindent
{\Large \bf Acknowledgements}\\
\noindent
The authors are indebted with Prof. F.Buccella for very useful
discussions and for his unfailing encouragement. We wish to express
our thanks also to Dr. P.Santorelli for very valuable talks and to S.De
Simone for his irreplaceable help for making figures.

\newpage

\begin{figure}[]
\caption[]{{\it
Charged-current Feynman diagram for $\nu_e$ self-energy in a
non-magnetized medium.}}
\end{figure}

\begin{figure}[]
\caption[]{{\it
Neutral-current Feynman diagram for $\nu_e$,$\nu_{\mu}$,$\nu_{\tau}$
self-energy in a non-magnetized medium.}}
\end{figure}

\begin{figure}[]
\caption[]{{\it
Charged-current Feynman diagram for $\nu_e$ self-energy in a
magnetized medium.}}
\end{figure}

\begin{figure}[]
\caption[]{{\it
Neutral-current Feynman diagram for $\nu_e$,$\nu_{\mu}$,$\nu_{\tau}$
self-energy in a magnetized medium.}}
\end{figure}


\begin{figure}[]
\caption[]{{\it
Level crossing diagram for $1 \, Mev$ momentum neutrino propagation
normal to the magnetic field
(this diagram coincides with that for $\vec{B} = 0$):
the continuous curves refer to $\omega_{1,2}$ in eq.(\ref{5.6}),
while the dashed lines to $\omega_{1,2}$ with zero mixing angle.
We use the typical values: $m_1 \simeq 10^{-8} \, eV$, $m_2 \simeq 10^{-3}
\, eV$, $sin^2 \, 2 \theta \simeq 7 \cdot 10^{-3}$.}}
\end{figure}


\begin{figure}[]
\caption[]{{\it
Level crossing diagram for $1 \, Mev$ momentum neutrino propagation
in a degenerate Fermi gas parallel to a magnetic field with strength $B
\simeq 10^{8} \, Gauss$. We use the same values of fig.5 for neutrino
masses and vacuum mixing angle.}}
\end{figure}


\begin{figure}[]
\caption[]{{\it
Level crossing diagram for $1 \, Mev$ momentum neutrino propagation
in a classical plasma parallel to a magnetic field with strength $B
\simeq 10^{14} \, Gauss$. We use the same values of fig.5 for neutrino
masses and vacuum mixing angle.}}
\end{figure}

\end{document}